\documentclass[preprintnumbers,amsmath,amssymbm,prd]{revtex4}
\usepackage{epsfig}
\usepackage{graphicx}
\usepackage{amssymb}

\begin{document}

\title{On the number of light rings in curved spacetimes of ultra-compact objects}
\author{Shahar Hod}
\affiliation{The Ruppin Academic Center, Emeq Hefer 40250, Israel}
\affiliation{ } \affiliation{The Hadassah Institute, Jerusalem
91010, Israel}
\date{\today}

\begin{abstract}

\ \ \ In a very interesting paper, Cunha, Berti, and Herdeiro have
recently claimed that ultra-compact objects, self-gravitating
horizonless solutions of the Einstein field equations which have a
light ring, must possess at least {\it two} (and, in general, an
even number of) light rings, of which the inner one is {\it stable}.
In the present compact paper we explicitly prove that, while this
intriguing theorem is generally true, there is an important
exception in the presence of degenerate light rings which, in the
spherically symmetric static case, are characterized by the simple
dimensionless relation $8\pi r^2_{\gamma}(\rho+p_{\text{T}})=1$
[here $r_{\gamma}$ is the radius of the light ring and
$\{\rho,p_{\text{T}}\}$ are respectively the energy density and
tangential pressure of the matter fields]. Ultra-compact objects
which belong to this unique family can have an {\it odd} number of
light rings. As a concrete example, we show that spherically
symmetric constant density stars with dimensionless compactness
$M/R=1/3$ possess only {\it one} light ring which, interestingly, is
shown to be {\it unstable}.
\end{abstract}
\bigskip
\maketitle

\section{Introduction}

One of the most physically interesting predictions of general
relativity is the existence of closed light rings in curved
spacetimes of compact astrophysical objects. These null circular
geodesics are usually associated with black-hole spacetimes (see
\cite{Chan,ShTe,CarC,Hodt1,Hodt2} and references therein), but
horizonless compact objects like boson stars may also possess light
rings \cite{CBH,Carm,Phi}.

Intriguingly, Cunha, Berti, and Herdeiro \cite{CBH} (see also
\cite{Carm}) have recently asserted that horizonless matter
configurations that have a light ring (the term ultra-compact
objects is usually used in the literature to describe these
self-gravitating field configurations) must have {\it pairs} (that
is, an even number) of light rings. In particular, the interesting
claim has been made \cite{CBH,Carm} that, for these regular
ultra-compact horizonless objects, one of the closed light rings is
{\it stable} \cite{Noteuns}.

Combining this claimed property of ultra-compact objects with the
intriguing suggestion made in \cite{Kei} (see also
\cite{CarC,Hodt1}) that, due to the fact that massless perturbation
fields tend to pile up on stable null geodesics, curved spacetimes
with stable light rings may develop non-linear instabilities, it has
been argued \cite{CBH,Carm,Hodt1} that horizonless ultra-compact
objects are non-linearly unstable and thus cannot provide physically
acceptable alternatives to the canonical black-hole solutions of the
Einstein field equations \cite{Notetnn}.

The main goal of the present paper is to prove that, while the
physically intriguing theorem presented in \cite{CBH} is generally
true, it may be violated by ultra-compact objects with degenerate
null circular geodesics which, in the spherically symmetric case,
are characterized by the simple dimensionless relation $8\pi
r^2_{\gamma}(\rho+p_{\text{T}})=1$ \cite{Noteexp}. In particular,
below we shall explicitly demonstrate that, in principle, one can
have special horizonless ultra-compact objects which possess only
{\it one} light ring which, interestingly, is shown to be {\it
unstable}.

\section{Description of the system}

We consider spherically symmetric nonlinear matter configurations
which are characterized by the static line element
\cite{Chan,ShTe,CarC,Hodt1,Hodt2,Noteunits,Notesc}
\begin{equation}\label{Eq1}
ds^2=-e^{-2\delta}\mu dt^2 +\mu^{-1}dr^2+r^2(d\theta^2 +\sin^2\theta
d\phi^2)\  ,
\end{equation}
where $\delta=\delta(r)$ and $\mu=\mu(r)$. Regular spacetimes are
characterized by the near-origin behavior \cite{Hodt1}
\begin{equation}\label{Eq2}
\mu(r\to 0)=1+O(r^2)\ \ \ \ {\text{and}}\ \ \ \ \delta(0)<\infty\  .
\end{equation}
In addition, asymptotically flat matter configurations are
characterized by the functional relations \cite{Hodt1}
\begin{equation}\label{Eq3}
\mu(r\to\infty) \to 1\ \ \ \ {\text{and}}\ \ \ \ \delta(r\to\infty)
\to 0\ .
\end{equation}

For spherically symmetric static spacetimes, the composed
Einstein-matter field equations, $G^{\mu}_{\nu}=8\pi T^{\mu}_{\nu}$,
are given by \cite{Hodt1,Noteprm}
\begin{equation}\label{Eq4}
\mu'=-8\pi r\rho+{{1-\mu}\over{r}}\
\end{equation}
and
\begin{equation}\label{Eq5}
\delta'=-{{4\pi r(\rho +p)}\over{\mu}}\  ,
\end{equation}
where $(\rho,p,p_{\text{T}})\equiv
(-T^{t}_{t},T^{r}_{r},T^{\theta}_{\theta}=T^{\phi}_{\phi})$ are
respectively the energy density, the radial pressure, and the
tangential pressure of the horizonless matter configuration
\cite{Bond1}. We shall assume that the matter fields satisfy the
dominant energy condition \cite{HawEl}
\begin{equation}\label{Eq6}
\rho\geq |p|,|p_T|\geq 0\  .
\end{equation}

The gravitational mass $m(r)$ of the field configuration contained
within a sphere of areal radius $r$ is given by the integral
relation \cite{Hodt1,Notemas}
\begin{equation}\label{Eq7}
m(r)=4\pi\int_{0}^{r} x^{2} \rho(x)dx\  .
\end{equation}
The relations (\ref{Eq6}) and (\ref{Eq7}) imply that regular matter
configurations with finite ADM masses (as measured by asymptotic
observers) are characterized by the asymptotic functional behavior
\begin{equation}\label{Eq8}
r^3p(r)\to 0\ \ \ \ \text{as}\ \ \ \ r\to\infty\  .
\end{equation}

Defining the dimensionless radial function
\begin{equation}\label{Eq9}
{\cal R}(r)\equiv 3\mu-1-8\pi r^2p\  ,
\end{equation}
one finds that the conservation equation
\begin{equation}\label{Eq10}
T^{\mu}_{r ;\mu}=0\
\end{equation}
together with the Einstein differential equations (\ref{Eq4}) and
(\ref{Eq5}) yield, for spherically symmetric static spacetimes, the
radial pressure gradient
\begin{eqnarray}\label{Eq11}
p'(r)= {{1} \over {2\mu r}}\big[{\cal R}(\rho+p)+2\mu T-8\mu p\big]\
,
\end{eqnarray}
where
\begin{equation}\label{Eq12}
T=-\rho+p+2p_T\
\end{equation}
is the trace of the energy-momentum tensor which characterizes the
self-gravitating matter fields.

\section{The number of light rings of spherically symmetric ultra-compact objects}

In the present section we shall explicitly prove that, while the
intriguing Cunha-Berti-Herdeiro theorem \cite{CBH} is generally
true, it may be violated by a special family of horizonless
ultra-compact objects that have a light ring which is characterized
by the dimensionless functional relation $8\pi
r^2_{\gamma}(\rho+p_{\text{T}})=1$.

To this end, we shall first derive, following the analysis of
\cite{Chan,CarC,Hodt1}, the characteristic functional relation of
null circular geodesics (light rings) in the spherically symmetric
static spacetime (\ref{Eq1}). Taking cognizance of the fact that the
curved line element (\ref{Eq1}) is independent of the time and
angular coordinates $\{t,\phi\}$, one deduces that the geodesic
trajectories are characterized by two conserved physical quantities:
the energy $E$ and the angular momentum $L$ \cite{Chan,CarC,Hodt1}.
In particular, the null geodesics of the spherically symmetric
static spacetime (\ref{Eq1}) are governed by an effective radial
potential $V_r$ which satisfies the characteristic equation
\cite{Chan,CarC,Hodt1,Notedot}
\begin{equation}\label{Eq13}
E^2-V_r\equiv \dot
r^2=\mu\Big({{E^2}\over{e^{-2\delta}\mu}}-{{L^2}\over{r^2}}\Big)\  .
\end{equation}

The characteristic circular geodesics of the horizonless curved
spacetime (\ref{Eq1}) are determined by the effective radial
potential (\ref{Eq13}) through the relations
\cite{Chan,CarC,Hodt1,Notethr}
\begin{equation}\label{Eq14}
V_r=E^2\ \ \ \ \text{and}\ \ \ \ V'_r=0\  .
\end{equation}
Taking cognizance of Eqs. (\ref{Eq4}), (\ref{Eq5}), and
(\ref{Eq13}), one can express (\ref{Eq14}) in the simple
dimensionless form \cite{Chan,CarC,Hodt1}
\begin{equation}\label{Eq15}
{\cal R}(r=r_{\gamma})=0\  .
\end{equation}
The remarkably compact functional relation (\ref{Eq15}) determines
the characteristic null circular geodesics of the spherically
symmetric static spacetime (\ref{Eq1}). For later purposes we note
that one learns from Eqs. (\ref{Eq2}), (\ref{Eq3}), and (\ref{Eq8})
that the dimensionless radial function ${\cal R}(r)$ [see Eq.
(\ref{Eq9})] is characterized by the simple relations
\begin{equation}\label{Eq16}
{\cal R}(r=0)=2\ \ \ \ \text{and}\ \ \ \ {\cal R}(r\to\infty)\to 2\
.
\end{equation}

As explicitly shown in \cite{Chan,CarC}, stable circular geodesics
of the spherically symmetric static spacetime (\ref{Eq1}) are
characterized by a locally convex effective radial potential with
the property $V''_r(r=r_{\gamma})>0$, whereas unstable circular
geodesics are characterized by a locally concave effective radial
potential with the opposite property $V''_r(r=r_{\gamma})<0$. Using
Eqs. (\ref{Eq4}), (\ref{Eq5}), (\ref{Eq9}) and (\ref{Eq11}), one
finds the compact functional relation
\begin{equation}\label{Eq17}
V''_r(r=r_{\gamma})=-{{E^2e^{2\delta}}\over{\mu r_{\gamma}}}\times
{\cal R}'(r=r_{\gamma})\
\end{equation}
for the second spatial derivative of the effective radial potential
(\ref{Eq13}), where [see Eqs. (\ref{Eq4}), (\ref{Eq9}), and
(\ref{Eq11})]
\begin{equation}\label{Eq18}
{\cal R}'(r=r_{\gamma})={{2}\over {r_{\gamma}}}\big[1-8\pi
r^2_{\gamma}(\rho+p_T)\big]\  .
\end{equation}

Taking cognizance of Eqs. (\ref{Eq15}) and (\ref{Eq16}), one deduces
that horizonless ultra-compact objects are generally (but, as will
be discussed below, not always) characterized by a discrete set
$\{r_{\gamma1},r_{\gamma2},...,r_{\gamma 2n-1},r_{\gamma 2n}\}$ of
{\it even} number of light rings with the property ${\cal
R}(r=r_{\gamma i})=0$. In particular, the subset
$\{r_{\gamma1},r_{\gamma3},...,r_{\gamma 2n-1}\}$ of odd light rings
is generally characterized by the property ${\cal R}'(r=r_{\gamma
i})<0$, whereas the subset $\{r_{\gamma2},r_{\gamma4},...,r_{\gamma
2n}\}$ of even light rings is generally characterized by the
property ${\cal R}'(r=r_{\gamma i})>0$. Thus, the first subset
corresponds to stable light rings with the property
$V''_r(r=r_{\gamma})>0$ [see Eq. (\ref{Eq17})], whereas the second
subset corresponds to unstable light rings with the property
$V''_r(r=r_{\gamma})<0$. It is important to stress the fact that
this conclusion agrees with the recently published interesting
theorem of Cunha, Berti, and Herdeiro \cite{CBH}.

However, it is also physically important to stress the fact that
there are special cases in which one (or more) of the light rings,
$r=r^*_{\gamma}$, which characterize the horizonless ultra-compact
objects is characterized by the functional relation \cite{Noterv}
\begin{equation}\label{Eq19}
{\cal R}'(r=r^*_{\gamma})=0\  .
\end{equation}
In this case the horizonless ultra-compact objects may be
characterized by an {\it odd} number of light rings.

In particular, as a special case that will be demonstrated
explicitly in the next section, we note that there are ultra-compact
objects which possess only {\it one} degenerate light ring. In
particular, this unique light ring is characterized by the simple
properties ${\cal R}(r=r^*_{\gamma})={\cal R}'(r=r^*_{\gamma})=0$
and ${\cal R}''(r=r^*_{\gamma})>0$. Interestingly, taking cognizance
of the characteristic properties $V_r(r\to0)\to\infty$ and
$V_r(r\to\infty)\to0$ of the effective radial potential (\ref{Eq13})
[see Eqs. (\ref{Eq2}) and (\ref{Eq3})] which determines the null
geodesics of the spacetime (\ref{Eq1}), one deduces that, in this
special case, the effective radial potential $V_r(r)$ is a
monotonically decreasing function with the simple property
$V_r(r=r^*_{\gamma}-\epsilon)>V_r(r=r^*_{\gamma})>V_r(r=r^*_{\gamma}+\epsilon)$
\cite{Noteeps}. Thus, the unique null circular geodesic
$r=r^*_{\gamma}$ \cite{Noteunc} is {\it unstable} to outward radial
perturbations. One therefore concludes that there are special cases
of horizonless ultra-compact objects which are characterized by the
relations (\ref{Eq15}) and (\ref{Eq19}) and which possess {\it no}
stable light rings.

\section{Ultra-compact objects with one light ring: constant density
stars}

In the present section we shall explicitly demonstrate that there
are ultra-compact horizonless objects which are characterized by an
odd number of light rings. In particular, we shall show that
constant density stars may possess only {\it one} light ring.

The Einstein-matter differential equations (\ref{Eq4}), (\ref{Eq5}),
and (\ref{Eq11}) can be solved analytically in the case of constant
density configurations, yielding the simple expression
\cite{ShTe,Carm,Notedp}
\begin{equation}\label{Eq20}
p(r)={{M}\over{{{4\pi}\over{3}}R^3}}\cdot{{\sqrt{3-{{6M}\over{R}}}-\sqrt{3-{{6Mr^2}\over{R^3}}}}\over
{\sqrt{3-{{6Mr^2}\over{R^3}}}-3\sqrt{3-{{6M}\over{R}}}}}\ \ \ ; \ \
\ 0\leq0\leq R\
\end{equation}
for the radial pressure, where $M$ and $R$ are respectively the
total mass and surface radius of the compact star \cite{Notepo}. In
addition, the exterior (vacuum) spacetime $r\geq R$ is described by
the Schwarzschild line element with total mass $M$, whereas the
interior spacetime of the compact star is characterized by the
metric component \cite{ShTe,Carm}
\begin{equation}\label{Eq21}
\mu(r)=1-{{2Mr^2}\over{R^3}}\ \ \ ; \ \ \ 0\leq0\leq R\  .
\end{equation}

Compact density stars with $M/R\geq1/3$ possess a light ring in the
exterior vacuum ($p=0$) region which, like the familiar
Schwarzschild black-hole spacetime, is characterized by the areal
radius
\begin{equation}\label{Eq22}
r_{\gamma}=3M\  .
\end{equation}
In addition, substituting Eqs. (\ref{Eq20}) and (\ref{Eq21}) into
the characteristic equation (\ref{Eq15}), which determines the null
circular geodesics of the spherically symmetric static spacetime
(\ref{Eq1}), one finds after some algebra that these compact objects
also possess a second (inner) light ring which is characterized by
the remarkably compact dimensionless relation
\begin{equation}\label{Eq23}
\mu(r_{\gamma})\cdot\Big(1-{{2M}\over{r_{\gamma}}}\Big)={{1}\over{9}}\
.
\end{equation}
Interestingly, from Eqs. (\ref{Eq21}), (\ref{Eq22}), and
(\ref{Eq23}) one deduces that constant density stars with $M/R=1/3$
are characterized by only {\it one} light ring with
$r_{\gamma}=R=3M$ \cite{Notedom,Notenwd,Notewlt,Bos}. These
horizonless matter configurations therefore violate the physically
interesting theorem presented in \cite{CBH,Notere}.

\section{Summary}

Horizonless ultra-compact objects with closed null circular
geodesics (light rings) have recently attracted much attention from
physicists and mathematicians (see \cite{CBH,Carm} and references
therein) as possible exotic alternatives to black-hole spacetimes
\cite{Noteuns}.

In a physically interesting paper, Cunha, Berti, and Herdeiro
\cite{CBH} have recently claimed that self-gravitating horizonless
solutions of the Einstein field equations which have a light ring,
must have at least {\it two} (and, in general, an even number of)
light rings, of which one of them is {\it stable}. Combined with the
non-linear instability to massless perturbation fields which is
expected to characterize curved spacetimes that possess stable light
rings \cite{Kei,CarC,Hodt1}, it has been claimed \cite{CBH,Carm}
that ultra-compact horizonless objects (self-gravitating field
configurations with light rings) cannot provide physically
acceptable alternatives to classical black-hole spacetimes
\cite{Notetnn,Noteos}.

In the present ultra-compact paper we have proved that, while the
physically important theorem presented in \cite{CBH} is generally
true, there is an intriguing exception for horizonless spacetimes
that possess degenerate null circular geodesics which, in the
spherically symmetric static case, are characterized by the simple
dimensionless functional relation [see Eqs. (\ref{Eq17}),
(\ref{Eq18}), and (\ref{Eq19})] \cite{Notens}
\begin{equation}\label{Eq24}
8\pi r^2_{\gamma}(\rho+p_{\text{T}})=1\  .
\end{equation}
We have stressed the fact that horizonless ultra-compact matter
configurations with the property (\ref{Eq24}) may have an {\it odd}
number of null circular geodesics. In particular, it has been
explicitly demonstrated that there are special cases of
ultra-compact objects which possess {\it no} stable light rings.

\bigskip
\noindent {\bf ACKNOWLEDGMENTS}

This research is supported by the Carmel Science Foundation. I would
like to thank Carlos A. R. Herdeiro, Emanuele Berti, and Pedro V. P.
Cunha for reading the manuscript and for their useful comments. I
would also like to thank Yael Oren, Arbel M. Ongo, Ayelet B. Lata,
and Alona B. Tea for stimulating discussions.


\begin{thebibliography}{99}
\bibitem{Noteos} It is important to stress the fact that, according to the
interesting theorem presented in \cite{CBH}, one of the
non-degenerate light rings of the created ultra-compact object has
to be stable if the matter content is physically reasonable (that
is, it respects the null energy condition \cite{CBH}).
\bibitem{Notetnn} It is important to note that this physically
interesting conclusion also depends on the characteristic timescale
of the expected non-linear instabilities \cite{CBH,Carm}.
\bibitem{Noteexp} Here $\{r_{\gamma},\rho,p_{\text{T}}\}$ are respectively
the radius of the light ring, the energy density, and the tangential
pressure of the matter fields.
\bibitem{Noteunits} We shall use natural units in which $G=c=1$.
\bibitem{Notesc} Here $(t,r,\theta,\phi)$ are the familiar
Schwarzschild spacetime coordinates.
\bibitem{Notethr} Note that these two characteristic conditions correspond to $\dot r^2=(\dot r^2)'=0$.
\bibitem{Noterv} Note that the relation (\ref{Eq19}) corresponds to a
special light ring with the property $V''_r(r=r^*_{\gamma})=0$ [see
Eq. (\ref{Eq17})].
\bibitem{Noteunc} It is worth emphasizing again that this special null circular geodesic is characterized
by the properties $V'_r(r=r^*_{\gamma})=V''_r(r=r^*_{\gamma})=0$.
\bibitem{Notedp} Note that the expression $3M/4\pi R^3$ in (\ref{Eq20}) is simply
the constant matter density of the star.
\bibitem{Notedom} It is worth noting that, for constant density stars, the dominant energy
condition $\rho\geq |p|,|p_T|\geq 0$ \cite{HawEl} is respected in
the physical regime $M/R\leq 3/8$ [see Eq. (\ref{Eq20})]. Thus, our
constant density stars with $M/R=1/3$ respect this energy condition.
\bibitem{Notenwd} Note that, for constant density stars, ${\cal R}'(r)$ is not well
defined at the surface $r=R$. This, however, does not harm our
observation that horizonless constant density stars with $M/R=1/3$
possess only one closed light ring.
\bibitem{Notere} It is worth emphasizing that the results of the present
paper reveal the intriguing fact that in some special (degenerate)
cases, one can have horizonless ultra-compact objects which posses
no stable null circular geodesics. In particular, as opposed to
generic ultra-compact objects which posses stable light rings, the
unstable degenerate light ring of these special ultra-compact matter
configurations is not expected to induce non-linear instabilities in
the corresponding spatially regular horizonless curved spacetimes.
Thus, one cannot rule out a priori the existence of stable
horizonless ultra-compact objects which possess degenerate light
rings. It would therefore be physically interesting to examine the
possibility of producing dynamically these ultra-compact objects
with degenerate null circular geodesics.
\bibitem{Notens} We believe that it would be physically interesting
to generalize the relation (\ref{Eq24}), which characterizes
ultra-compact objects that can violate the important theorem
presented in \cite{CBH}, to the non-spherically symmetric regime of
self-gravitating horizonless matter configurations.
\end{thebibliography}
\end{document}